\documentclass[fleqn,usenatbib]{mnras}

\usepackage{newtxtext,newtxmath}

\usepackage[T1]{fontenc}

\DeclareRobustCommand{\VAN}[3]{#2}
\let\VANthebibliography\thebibliography
\def\thebibliography{\DeclareRobustCommand{\VAN}[3]{##3}\VANthebibliography}

\usepackage{graphicx}	
\usepackage{amsmath}	

\title[Commensal Searches in SGRB Fields]{Commensal Transient Searches in Eight Short Gamma Ray Burst Fields}

\author[S. I. Chastain et al.]{
S. I. Chastain,$^{1}$\thanks{E-mail: schastain@gwu.edu (SIC)}
A. J. van der Horst,$^{1}$
A. Rowlinson,$^{2,3}$
L. Rhodes,$^{4}$
A. Andersson,$^{5}$
R. Diretse,$^{6}$
\newauthor
R. P. Fender,$^{5}$
and P. A. Woudt$^{7}$
\\
$^{1}$Department of Physics, George Washington University, 725 21st St NW, Washington, DC, 20052, USA\\
$^{2}$Anton Pannekoek Institute, University of Amsterdam, Postbus 94249, 1090 GE Amsterdam, The Netherlands\\
$^{3}$ASTRON, the Netherlands Institute for Radio Astronomy, Oude Hoogeveensedijk 4, 7991 PD, Dwingeloo, The Netherlands\\
$^{4}$Astrophysics, Department of Physics, University of Oxford, Keble Road, Oxford, OX1 3RH, UK \\
$^{5}$Department of Physics, Astrophysics, University of Oxford, Denys Wilkinson Building, Keble Road, Oxford OX1 3RH, UK \\
$^{6}$Inter-University Institute for Data-Intensive Astronomy, Department of Astronomy, University of Cape Town, Private Bag X3, Rondebosch 7701, South Africa. \\ 
$^{7}$Department of Astronomy, University of Cape Town, Private Bag X3, Rondebosch 7701, South Africa
}

\date{Accepted XXX. Received YYY; in original form ZZZ}

\pubyear{2022}

\begin{document}
\label{firstpage}
\pagerange{\pageref{firstpage}--\pageref{lastpage}}
\maketitle

\begin{abstract}
A new generation of radio telescopes with excellent sensitivity, instantaneous {\it uv} coverage, and large fields of view, are providing unprecedented opportunities for performing commensal transient searches. Here we present such a commensal search in deep observations of short gamma-ray burst fields carried out with the MeerKAT radio telescope in South Africa at 1.3 GHz. These four hour observations of eight different fields span survey lengths of weeks to months. We also carry out transient searches in time slices of the full observations, at timescales of 15 minutes, and 8 seconds. We find 122 variable sources on the long timescales, of which 52 are likely active galactic nuclei, but there are likely also some radio flaring stars. While the variability is intrinsic in at least two cases, most of it is consistent with interstellar scintillation. In this study, we also place constraints on transient rates based on state-of-the-art transient simulations codes. We place an upper limit of $2\times10^{-4}$ transients per day per square degree for transients with peak flux of 5 mJy, and an upper limit of $2.5\times10^{-2}$ transients per day per square degree for transients with a fluence of 10 Jy ms, the minimum detectable fluence of our survey.
\end{abstract}




\section{Introduction}
Transient searches at radio wavelengths are now yielding an unprecedented number of transients of all kinds. For some time now, transient searches in other wavebands such as the optical and X-rays have yielded a large number of results, and now this is starting to be true for the radio regime as well. Some searches in time series analysis have found transients like fast radio bursts (FRBs) with timescales on the order of milliseconds~\citep[e.g.,][]{2007Sci...318..777L,2021ApJS..257...59C}. Other searches have been performed in radio images, with the number of transients and variables found this way increasing and yielding interesting results. For example, a transient was found in the LOFAR Multi-frequency Snapshot Survey on a timescale of around 10 minutes at 60 MHz~\citep{2016MNRAS.456.2321S}. The CHILES Variable and Explosive Radio Dynamic Evolution Survey~\citep{2021ApJ...923...31S} spent hundreds of hours observing the COSMOS field at 1.4 GHz and found a number of variable sources at timescales from days to years. There have also been transients found as part of a commensal search, that is, a search of data taken as part of a different scientific objective. In commensal transient searches with MeerKAT at 1.3 GHz,~\citet{2020MNRAS.491..560D} find a transient with a timescale of weeks and a variable pulsar on sub-week timescales; and in this same field,~\citet{2022MNRAS.512.5037D} find variable sources on timescales of weeks to months. Similarly,~\citet{2022MNRAS.517.2894R} found four variable sources with timescales spanning from seconds up to over a year. \citet{2022MNRAS.513.3482A} also found a radio transient source in a commensal search of MeerKAT data. As part of Deeper, Wider, Faster,~\citet{2023MNRAS.519.4684D} have found multiple transients and variables with ASKAP. Additionally, the Variables and Slow Transients survey (VAST) using ASKAP~\citep{2021PASA...38...54M} has found multiple radio transients~\citep{2021ApJ...920...45W,2022MNRAS.516.5972W} and the Very Large Array Sky Survey (VLASS) using the VLA at frequencies around 3 GHz~\citep{2020PASP..132c5001L} promises to find a large number of transients and variables due to their large sky coverage and multi-epoch observing strategies. 

Commensal searches for transients and variables is proving to be a valuable way of probing the radio sky, in particular with facilities that have a large field of view. Not only are commensal searches an efficient use of pre-existing scientific data, they also have the potential to find new and interesting sources as well as increasing our knowledge of the populations of sources on the radio sky by constraining transient rates~\citep[e.g.,][]{2011ApJ...728L..14B,2016MNRAS.459.3161C}. The number of detections along with the survey properties, if used in conjunction with accurate transient rate calculations, can uncover more information about sources with unknown associations. In addition, with calculations that allow for calculating different transient rates for different parts of the sky, such as~\citep{2022ascl.soft04007C}, it is possible to reveal differences in transients and their behavior in different parts of the sky, such as galactic versus extragalactic sources. 

Enabling all of these aforementioned new transient discoveries, with their excellent sensitivity and large field of view, are new facilities such as MeerKAT, ASKAP, and LOFAR~\citep{2009IEEEP..97.1522J,2008ExA....22..151J,2013A&A...556A...2V}. Due to the excellent instantaneous {\it uv} coverage of these instruments, these searches are also able to probe increasingly shorter timescales, with the capability to image on timescales down to seconds, or to create deep images that combine many hours' worth of data. All of these improvements are creating a wealth of new opportunities for commensal transient searches in radio images. 

ThunderKAT~\citep{2016mks..confE..13F} is a large survey project for image plane radio transients with MeerKAT. Taking advantage of the new opportunities provided by MeerKAT is a key part of its mission, as it includes conducting commensal transient searches in MeerKAT imaging data (besides performing follow-up observations of specific transients found in other wavebands). Part of the challenge of these searches are that it requires analyzing a large amount of data, on the order of hundreds of gigabytes to terabytes. In order to search through these images, we use the LOFAR transients pipeline \citep[{\sc TraP};][]{2015A&C....11...25S}, which creates a catalog of sources and their light curves, and tracks the variability of all the sources in the images. Using TraP we conduct a commensal transient search on multiple timescales of short gamma-ray burst observations taken as part of the ThunderKAT project. We establish methodologies and techniques to find new variable sources among the large quantity of sources in this dataset. We also look into whether the variability of these sources is intrinsic or extrinsic (e.g., interstellar scintillation), and draw conclusions to guide future similar studies.

We will describe the observations and overall data set in Section~2, and the methodology for the transient search in Section~3. The results are presented in Section~4 and discussed in Section~5, with a summary and concluding remarks in Section~6.

\section{Observations}

We performed a commensal transient search in observations of eight short gamma-ray burst fields. These fields represent all short GRB fields that were observed in the first three years of the ThunderKAT project. Each observation was about 4 hours in duration including overhead, such as calibrator observations, with the number of observations per field varying. Each 4-hour observation consisted of 15-minute scans with calibrator measurements of a few minutes interspersed. Table~\ref{tab:allobs} lists all the observations that are a part of our transient survey. Since our search radius was 0.8 degrees from the center of each image and we had 8 different fields, the total survey area is 16.1 square degrees. After doing some quality control for bright sources, our survey area is reduced to 16.0 square degrees. The observations were calibrated using version 1.1 of the ProcessMeerKAT pipeline~\citep[{\sc ProcessMeerKAT};][]{pminprep}. As part of this calibration process, parts of the spectrum with a large amount of known radio frequency interference (RFI) were flagged, resulting in a bandwidth of about 800 MHz centered at 1.28 GHz. Calibration was performed in parallel by separating the measurement set into 11 spectral windows. The bandpass and complex gain calibration was performed using Common Astronomy Software Applications \citep[{\sc CASA;}][]{2022arXiv221002276T} tools and the calibrators listed in Table~\ref{tab:allobs}. Automated RFI flagging was performed using tfcrop and rflag. After two rounds of calibration and flagging, the spectral windows were recombined into a single measurement set for imaging. 

All images were made using tclean with the initial image being at least 5120 pixels of 2 arcseconds in size, in order to include the entire primary beam in the image. The 4-hour images were made by producing a shallow image with the cleaning process stopping based on a threshold of 1 mJy; and then self-calibration and flagging for RFI was performed before making the final, deep image stopping at a threshold of around 80 $\mu$Jy. The 15-minute images were made using the self-calibrated measurement set. The imaging parameters used include the multi-term multi-scale imaging algorithm with the w-project gridder. For the 4-hour images, 128 w-planes were used; and for the 15-minute images, 64 w-planes were used. The latter resulted in increased correlated noise in the 15-minute images. 

The shortest imaging timescale of the data is determined by the integration time of the observations, which is 8 seconds for every observation in this survey. On this timescale the imaging parameters were slightly different. The quality of images made using w-projection and those not using w-projection were seen to be quite similar, apart from a slight offset in the spatial coordinates between the two. Therefore, in an effort to save processing time and computational resources, the images were made using the standard gridder without w-projection, using the multi-term multi-scale imaging algorithm that is a part of tclean.

\subsection{Image Quality}

\begin{figure}
    \includegraphics[width=\columnwidth]{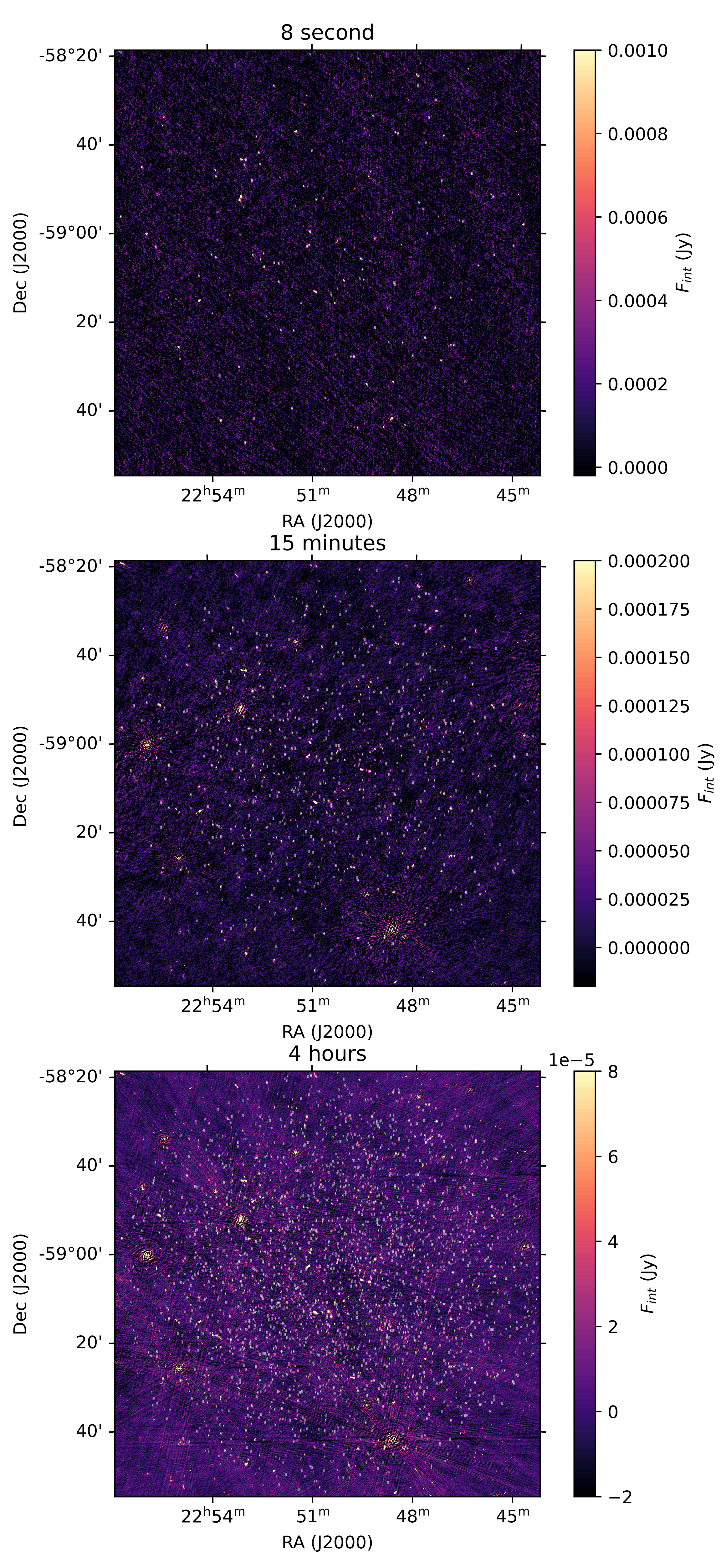}
    \caption{Example images from each of the three timescales. The 8 second image has an rms noise of 176 $\mu$Jy, the 15 minute image has an rms noise of 25 $\mu$Jy, and the the 4 hour image has an rms noise of 11 $\mu$Jy. Marked with white boxes are all the sources at or above the five sigma level detected by TraP within 0.8 degrees of the center of the image after excluding noisy regions and only allowing one source within a radius of five beamwidths of any other source.}
    \label{fig:allsrcs}
\end{figure}

The typical noise values roughly follow the expected scaling for noise as a function of observation time $t$, that is $~1/\sqrt(t)$, and are summarized in Table~\ref{tab:noisesummary} below. As the timescales go shorter, the trend is for the variance in the noise values to go larger, with the images on the 8-second timescale showing a large range of values. 31 out of the 47,964 images at the 8-second timescale had a noise that was many orders of magnitude higher than the typical noise distribution, skewing the statistics in a way that is possibly misleading, and therefore we exclude these highest 31 values for noise on this timescale. In the actual analysis we did not perform any additional quality control steps in the version of TraP we used. Figure~\ref{fig:allsrcs} shows an example image of each timescale with all of the sources detected at the five sigma level or greater by TraP.

\begin{table*}
\caption{All observations used in this study, indicating the observations' start and end times, phase center position, time spent on the target, and calibrators used. Each observation was at least 2.8 degrees by 2.8 degrees.}
\label{tab:allobs}
\begin{tabular}{lllllll}
\hline
       Name  &    Observation Time &         RA  &        DEC &  Time On Target (hrs)  & Bandpass Calibrator & Gain Calibrator\\
\hline
GRB 200219A &  2020-02-19T14:13:47 to 18:16:47 & 342.6385 & -59.1196 &              2.6189 & J0408-6545 & J2329-4730 \\
 GRB 200219A &  2020-02-21T12:28:44 to 16:28:44 & 342.6385 & -59.1196 &              3.2031 & J0408-6545 & J2329-4730 \\
 GRB 200219A &  2020-02-23T12:07:21 to 16:07:21 & 342.6385 & -59.1196 &              3.1986 & J0408-6545 & J2329-4730 \\
 GRB 200219A &  2020-02-27T13:52:43 to 17:51:43 & 342.6385 & -59.1196 &              3.1986 & J0408-6545 & J2329-4730 \\
 GRB 200411A &  2020-04-12T07:07:12 to 11:09:12 &  47.6641 & -52.3176 &              3.2053 & J0408-6545 & J0210-5101 \\
 GRB 200411A &  2020-04-14T11:31:27 to 15:32:27 &  47.6641 & -52.3176 &              3.2031 & J0408-6545 & J0210-5101 \\
Sculptor & 2020-04-16T05:15:42 to 09:17:42 &  11.8875 & -25.2886 &              3.2031 & J1939-6342 & J0025-2602 \\
GRB 200411A & 2020-04-18T07:27:35 to 11:29:35 &  47.6641 & -52.3176 &              3.2009 & J0408-6545 & J0210-5101 \\
GRB 200522A & 2020-05-23T06:56:37 to 11:10:37 &   5.6820 &  -0.2832 &              3.4496 & J1939-6342 & J0022+0014 \\
GRB 200522A & 2020-05-24T06:01:09 to 10:15:09 &   5.6820 &  -0.2832 &              3.4496 & J1939-6342 & J0022+0014 \\
GRB 200522A & 2020-05-29T02:11:13 to 06:24:13 &   5.6820 &  -0.2832 &              3.4452 & J1939-6342 & J0022+0014 \\
GRB 200522A & 2020-06-06T02:01:14 to 06:15:44 &   5.6820 &  -0.2832 &              3.4474 & J1939-6342 & J0022+0014 \\
GRB 200907B & 2020-09-08T01:03:47 to 05:23:17 &  89.0290 &   6.9062 &              3.4430 & J0408-6545 & J0521+1638 \\
GRB 200907B & 2020-09-10T01:47:12 to 06:05:42 &  89.0290 &   6.9062 &              3.4541 & J0408-6545 & J0521+1638 \\
GRB 200907B & 2020-09-14T01:35:52 to 05:54:16 &  89.0290 &   6.9062 &              3.4585 & J0408-6545 & J0521+1638 \\
GRB 200907B & 2020-09-25T02:17:12 to 06:34:54 &  89.0290 &   6.9062 &              3.4563 & J0408-6545 & J0521+1638 \\
GRB 210323A & 2021-03-25T06:17:56 to 10:37:26 & 317.9461 &  25.3699 &              3.4519 & J1939-6342 & J2236+2828 \\
GRB 210323A & 2021-03-27T06:03:55 to 10:23:16 & 317.9461 &  25.3699 &              3.4519 & J1939-6342 & J2236+2828 \\
GRB 210323A & 2021-04-01T05:37:48 to 09:57:17 & 317.9461 &  25.3699 &              3.4563 & J1939-6342 & J2236+2828 \\
GRB 210726A & 2021-07-28T14:24:49 to 17:50:28 & 193.2909 &  19.1875 &              2.7122 & J1331+3030 & J1330+2509 \\
GRB 210726A & 2021-08-01T12:28:16 to 16:47:14 & 193.2909 &  19.1875 &              3.4519 & J1331+3030 & J1330+2509 \\
GRB 210726A & 2021-08-07T12:07:14 to 16:26:20 & 193.2909 &  19.1875 &              3.4519 & J1331+3030 & J1330+2509 \\
GRB 210726A & 2021-08-19T12:18:07 to 16:36:33 & 193.2909 &  19.1875 &              3.4519 & J1331+3030 & J1330+2509 \\
GRB 210726A & 2021-09-06T11:38:11 to 15:56:29 & 193.2909 &  19.1875 &              3.4496 & J1331+3030 & J1330+2509 \\
GRB 210919A & 2021-09-20T01:22:10 to 05:40:20 &  80.2545 &   1.3115 &              3.4519 & J0408-6545 & J0503+0203 \\
GRB 210919A & 2021-09-24T02:49:58 to 07:08:40 &  80.2545 &   1.3115 &              3.4541 & J0408-6545 & J0503+0203 \\
GRB 210726A & 2021-09-26T09:35:20 to 13:54:18 & 193.2909 &  19.1875 &              3.4563 & J1331+3030 & J1330+2509 \\
GRB 210919A & 2021-09-27T01:23:09 to 05:41:27 &  80.2545 &   1.3115 &              3.4541 & J0408-6545 & J0503+0203 \\
GRB 210323A & 2021-09-30T17:35:55 to 21:56:06 & 317.9461 &  25.3699 &              3.4541 & J1939-6342 & J2236+2828 \\
GRB 210726A & 2021-12-27T02:48:11 to 07:07:17 & 193.2909 &  19.1875 &              3.4430 & J1331+3030 & J1330+2509 \\
\end{tabular}	
\end{table*}

\begin{table}
\caption{Summary of the mean, median and range of the image noise distributions at each timescale in our study. Note that the 8-second timescale statistics are computed with the highest 31 noise values excluded.}
\label{tab:noisesummary}
\begin{tabular}{llll}
\hline
Timescale & Range ($\mu$Jy) & Median ($\mu$Jy)   & Mean ($\mu$Jy)  \\
 \hline
4 hours & 6 to 32 & 10 & 13 \\
15 minutes & 19 to 184 & 30 & 43 \\
8 seconds & 106 to 17709 & 176 & 205 \\ 
\end{tabular}	
\end{table}

\begin{table}
\caption{Summary of number of images of each field at each timescale.}
\label{tab:obstimescales}
\begin{tabular}{l|c|c|c}
\hline
Target & 4 hour images & 15 minute images & 8 second images \\
\hline
GRB200219A &	4&	51&	5552\\
GRB200411A &	3&	39&	4365\\
Sculptor	&1	&13&	1455\\
GRB200522A	&4&	54&	6265\\
GRB200907B&	4&	56&	6274\\
GRB210323A&	4&	56&	6275\\
GRB210726A&	7&	95&	9072\\
GRB210919A&	3&	42&	4706\\
\end{tabular}
\end{table}

\section{Methods}

\subsection{Transient Searches with the LOFAR Transients Pipeline}

After calibrating the data and producing images, the latter were run through the TraP~\citep{2015A&C....11...25S} version 4. While originally designed for LOFAR, the TraP is telescope-agnostic and well suited for any kind of image based radio transient search. When running the images through the pipeline, a detection threshold of 5$\sigma$ was used, which is the threshold for blind detection of a source, along with an analysis threshold of $3\sigma$, which is the threshold used for analyzing information about the source (such as position and flux, and uncertainties in those quantities). The detection threshold was set to 5$\sigma$ instead of the default 8$\sigma$ so that more sources would be detected and analyzed by TraP. We later increase this threshold and reduce the number of candidate transients and variables through additional analysis, as described in section ``Determining Candidate Variables and Transients.'' This process of starting with a lower threshold and increasing it later was beneficial for capturing longer portions of variable light curves since, when a variable source reaches the detection threshold in TraP, the TraP does not go back to previous images to measure the flux of the source before detection. The TraP calculates variability statistics $V$ and $\eta$, which the user can use to classify a source as constant or varying. These statistics are as defined in \citet{2015A&C....11...25S} where $I$ is the flux measurement of a source; $\xi$ is the average flux weighted by the inverse of the flux measurement errors, $\sigma$; and averages are indicated by hyphens above the quantities in these equations:
\begin{equation}\label{Veqn}
    V_{\nu} = \frac{1}{\bar{I}_{\nu}}\sqrt{\frac{N}{N-1} (\bar{I_{\nu}^2} - \bar{I_{\nu}}^2)}
\end{equation}
\begin{equation}\label{etaeqn}
    \eta =  \frac{1}{N-1} \sum_{i=1}^{N} \frac{(I_{\nu,i} - \xi_{I_{\nu}})^2}{\sigma_{\nu,i}^2}   
\end{equation}
The transient search on the 8-second timescale was limited in time to the images contained within an approximately 15 minute scan in order to improve variability statistics and due to the warping of the coordinates in the images. Furthermore, the beamwidths limit in the TraP parameters was set to 3, which relaxes the hard limits on the association between sources that are spatially separated. After searching each of these 15 minutes for transients, the images containing transients were then re-imaged using w-projection, and then compared with the previous images to acquire a corrected position. The number of images for each field is shown in Table~\ref{tab:obstimescales}. Initial runs used 10 deblending thresholds, which is intended to separate sources that are very close together into separate sources; however, due to errors involving source identification and the database within TraP in which an extremely large number of sources were located at the same coordinates in the images, later runs used zero deblending thresholds. This change does not affect any potential transient or variable sources, since these sources would have been clustered close together and discarded in the next step when we restrict the number of sources within 5 times the major axis of the beam (see below).

The output from TraP contains a large number of sources, many of which are not astrophysical but features resulting from imaging artifacts such as sidelobes. These imaging artifacts tend to show up as patterns of bright and dark spots around a relatively bright source. In order to eliminate these sources, sources within a region of a radius of 5 times the major axis of the point-spread function of the brightest sources are discarded. The radius, in beamwidths, was determined through some trial-and-error, and in future studies can be increased to reduce artifacts detected as transients or decreased to reduce the chance of eliminating sources that happen to be tightly clustered in the image. The deep 4-hour images were also examined for each field, and regions excluded from the transient search were created around areas of poor quality due to extremely bright sources. The total area of these excluded regions were 0.11 square degrees and the lowest flux of a source that was excluded was around 12 mJy.

\subsection{Determining Candidate Variables and Transients}
\label{sec:determinecand}

One challenge of performing transient searches is deciding on the appropriate signal-to-noise cut for a source detection. For this study we adopt the methodology used in~\citet{2022MNRAS.517.2894R}: we fit the flux values of all the pixels in the images to a Gaussian, to determine the sigma threshold that would result in less than one false positive. In the case of the 15-minute and 8-second images, we use a sub-set of the images as a sample, 50\% and 2.5\% of the images, respectively. These sample sizes were constrained by the size of memory of the machine used to compute the threshold. In order to scale up the calculated thresholds to account for the images that were not selected, we used a scaling factor to scale up to the number of pixels that would be in the entire dataset. We then used the quantile function of the Gaussian distribution to determine the sigma threshold that would result in less than one false positive. We did this for all fields combined together but for every timescale separately, and found that the thresholds are approximately 5.3$\sigma$, 5.7$\sigma$, and 6.4$\sigma$ for the 4-hour, 15-minute, and 8-second images respectively. We then reduced the number of potentially interesting sources to investigate by making cuts based on these sigma thresholds, excluded regions, and proximity to neighboring sources.

\begin{figure*}
    \includegraphics[width=0.9\textwidth]{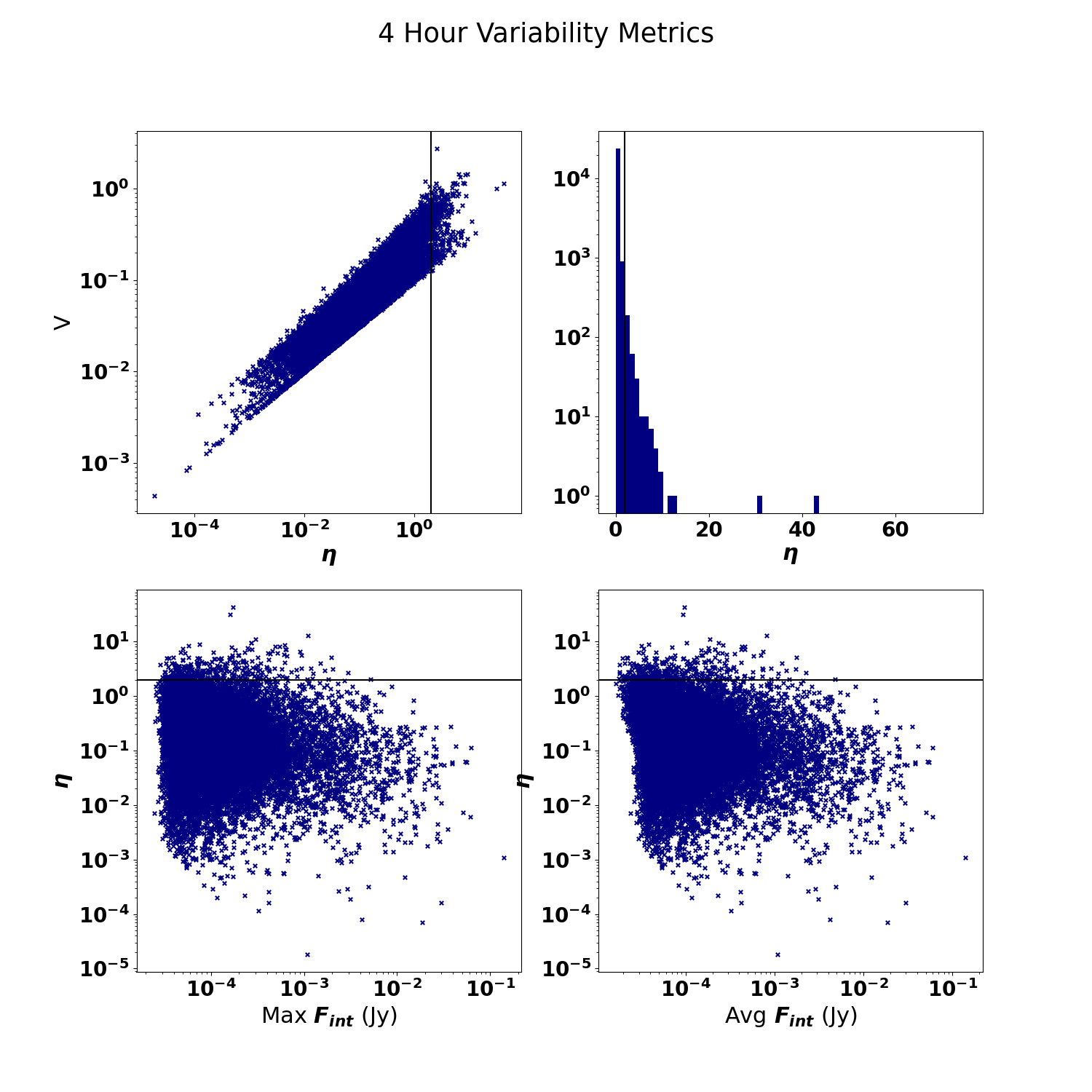}
    \caption{Variability statistics $V$ and $\eta$, as defined in equations~\ref{Veqn} and~\ref{etaeqn}, for the 4-hour timescale, also versus the maximum and average integrated flux. A black line vertical line marks $\eta=2$.}
    \label{fig:varstat4hr}
\end{figure*}

\begin{figure*}
    \includegraphics[width=0.9\textwidth]{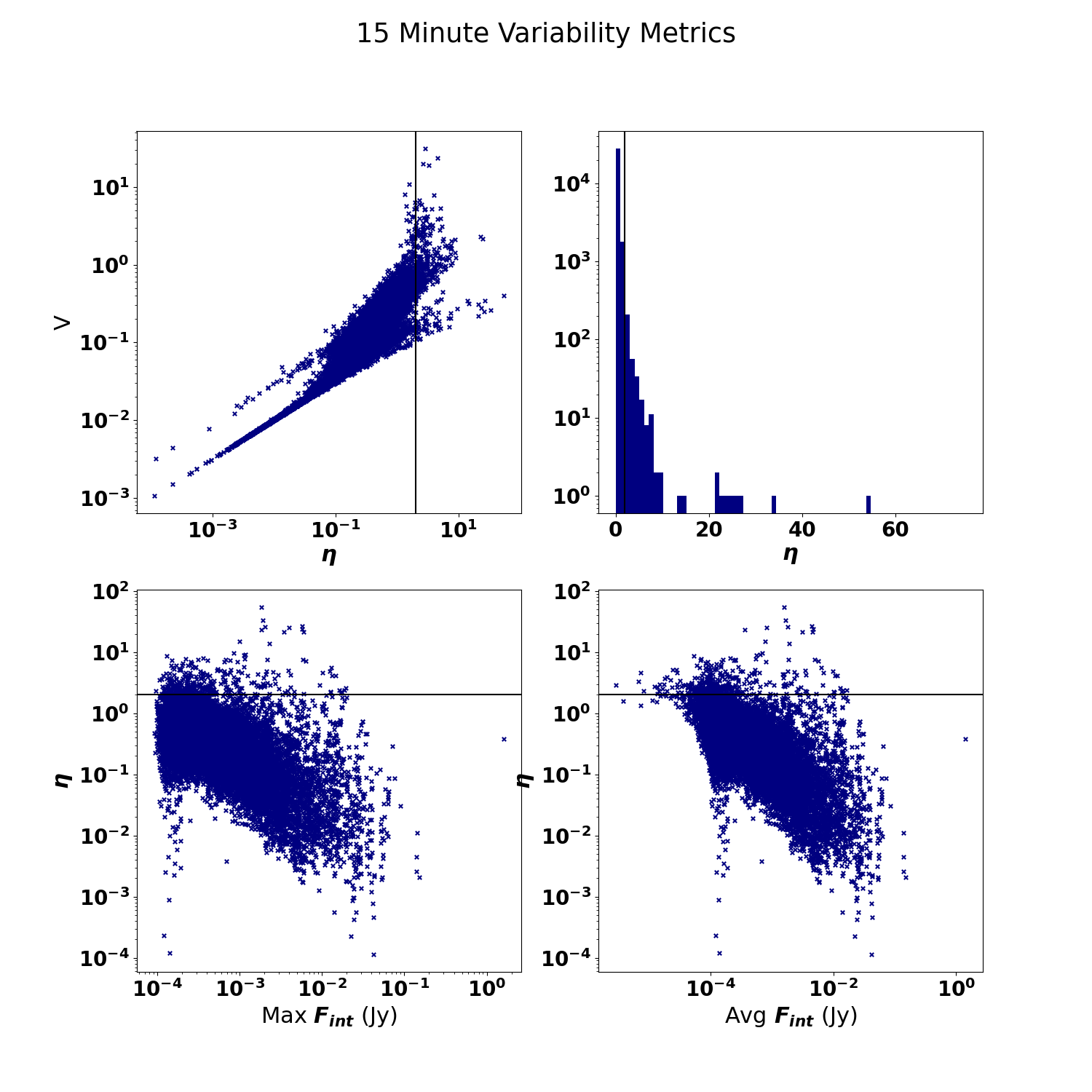}
    \caption{Variability statistics $V$ and $\eta$, equations~\ref{Veqn} and~\ref{etaeqn}, for the 15-minute timescale, also versus the maximum and average integrated flux. A black line vertical line marks $\eta=2$.}
    \label{fig:varstat15min}
\end{figure*}

\begin{figure*}
    \includegraphics[width=0.9\textwidth]{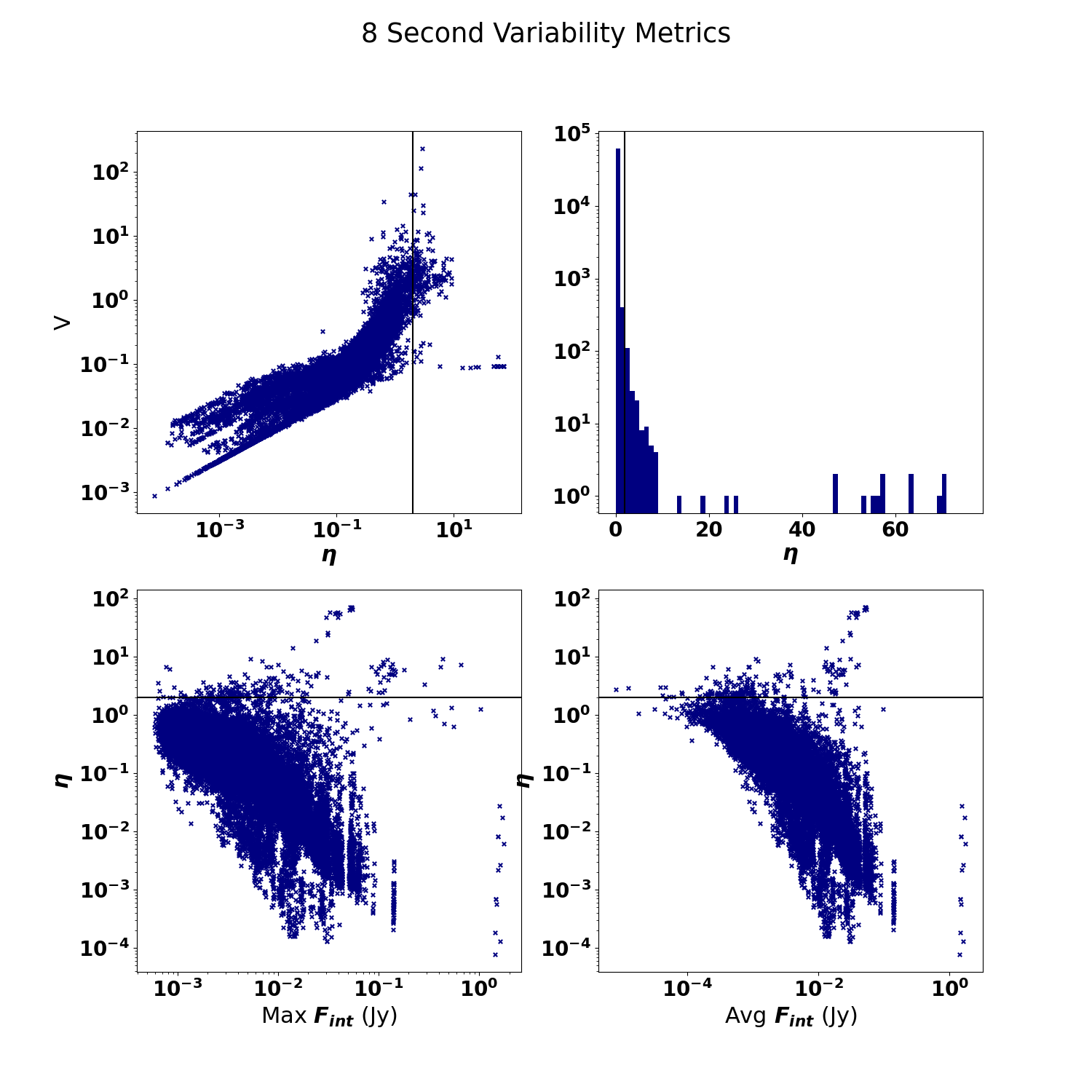}
    \caption{Variability statistics $V$ and $\eta$, as defined in equations~\ref{Veqn} and~\ref{etaeqn}, for the 8-second timescale, also versus the maximum and average integrated flux. A black line vertical line marks $\eta=2$.}
    \label{fig:varstat8sec}
\end{figure*}

After removing potential transients that fall below the signal-to-noise thresholds determined above, in order to determine which sources are potentially variable or transient, we recalculate the variability statistic $\eta$ accounting for an assumed $10\%$ systematic error, since the value that the TraP calculates for $\eta$ does not include systematic errors that can arise due to instrumental effects and/or calibration errors. We chose $10\%$ for systematic error to account for additional uncertainty due to calibration errors and the observations being taken in the first few years of MeerKAT. We then create animations to examine all sources with a corrected $\eta$ above a value of 2. This value was chosen for practicality reasons, since it represents a value of $\eta$ that indicates significant variability and results in a number of sources that could be practically examined in a non-automated way. This resulted in 214 sources in the 8-second images, 306 sources in the 15-minute images, and 278 sources in the 4-hour images. The images of the sources over time were turned into the aforementioned animations with light curves and variability parameters plotted\footnote{Code available at https://github.com/dentalfloss1/sharedscripts as FindOutliers.py}. Using these animations along with light curves, the sources were sorted by eye into one of the following categories: potentially interesting astrophysical transients, obvious noise artifacts, misassociation errors, and moving objects. The moving objects were only found on the 8-second timescale and are most likely due to RFI, due to the narrow-band behavior of the few sources bright enough for spectral analysis. The objects classified as noise artifacts were mostly from large noise patterns across the field that were mistakenly detected as sources. The misassociation issues came from a number of sources, but most of them due to sidelobes around bright sources.

\section{Results}
\label{sec:results2}
After examining the sources by eye, the number of potential astrophysical transients was three in the 8-second images, 19 in the 15-minute images, and 227 in the 4-hour images. To examine the potential transients on the 8-second timescale, corrected positions were acquired by making w-projected images of all of the 8-second integrations that made up the two scans closest to the time in which the transient appears. Then, a second run through the TraP was done with a forced fit at the corrected position of the transient location. After this process, due to changes in the noise from w-projection, all three potential transients fell below the detection threshold of 6.4$\sigma$ that we previously determined. We followed a similar process for the 15 minute images, forcing a fit at the location of the transients on the 15-minute timescale. After this process, we recalculated the corrected $\eta$ for these sources and no candidates remained at this timescale. Plots showing the flux and variability statistics of the sources on the three different timescales are shown in Figures~\ref{fig:varstat4hr}-\ref{fig:varstat8sec}.

All of the candidates in the 4 hour images were variables. To ensure that these variations were significant, we used the katbeam library to correct for the sensitivity of the primary beam~\citep{2022AJ....163..135D}. We then once again did a forced fit at all of the sources' locations. As a result of the force fitting and once again recalculating a corrected $\eta$, we find 122 sources that still have a corrected $\eta$ greater than two. These sources are all considered to be candidate variables. 

\subsection{Matching Catalogs}
\label{sec:MatchingCatalogs}
In order to better understand the variable sources, a search was performed of catalogs available within Vizier~\citep[]{vizier} using the Astroquery python library~\citep{2019AJ....157...98G}. These catalogs include other radio catalogs such as FIRST, VLASS,  and NVSS \citep{2015ApJ...801...26H,2021ApJS..255...30G,1998AJ....115.1693C}, in addition to a large number of optical, infrared, and near-infrared catalogs, and some X-ray and gamma-ray catalogs. Notably some of the fields in this survey lack significant multi-wavelength observations due to a lack of surveys from observatories in the southern hemisphere. In addition to searching Vizier, we also searched the Living Swift XPS catalog~\citep{2022MNRAS.tmp.2790E} for X-ray counterparts. The closest LSXPS source to any of our variable sources was approximately 65 arcseconds away, and therefore we conclude that there are no matches in this catalog. For the other catalogs, of the 122 variables in the 4-hour images, 100 of them have a source in other catalogs that are within one arcsecond, which we consider a catalog match. For these catalog matches, a false association probability of 0.05 was determined in a similar manner to ~\citet{2015ApJ...801...26H}, by offsetting the source positions by one arcminute and testing to see how many sources have catalog counterparts. There are 22 sources with no catalog matches, which are given in Table~\ref{tab:unmatched}. 17 of these sources are at southern declinations where there is a lack of catalog data. However, five of these sources should be visible to many different facilities. The lack of a catalog match to these sources could be due to properties of the source, such as the spectral index or intervening material, and the lack of matches in radio catalogs could be due to their relatively low observed flux level on the order of hundreds of $\mu$Jy. To investigate these sources further, a forced flux measurement was performed using TraP at the locations of each of these five sources in the 1.1, 2.1, and 3.1 epochs of the VLASS quick-look data. Three of the five sources had force flux measurements at or below the MeerKAT flux measurements and a 3$\sigma$ limit greater than the MeerKAT flux measurements as well. Sources 713985 and 714807 showed marginal source detections.

\begin{table}
\caption{22 variable sources with no multi-wavelength counterparts along with their positions and average integrated flux measurements from MeerKAT images. }
\begin{tabular}{c|c|c|c|c} 
\hline
id &       RA &    Dec   & Field & $F_{avg,int}$ ($\mu$Jy)\\
\hline
 693964 &  344.1593 & -59.5321 & GRB 200219A &43\\
 694324 &  343.5354 & -58.8087   & GRB 200219A &133\\
 694542 &  343.3390 & -58.9357 & GRB 200219A &50\\
 695276 &  342.8254 & -58.5346 & GRB 200219A &193\\
 696983 &  341.6460 & -59.6791 & GRB 200219A &131\\
713623 &   89.7308 &   6.6844   & GRB 200907B &60\\
713985 &   89.5435 &   6.3984   & GRB 200907B &72\\
714170 &   89.4624 &   6.9762   & GRB 200907B &116\\
 714807 &   89.2317 &   7.0995 & GRB 200907B &457\\
716711 &   88.4060 &   6.6692   & GRB 200907B &77\\
 702092 &   48.5602 & -52.5190 & GRB 200411A &28\\
 702209 &   48.4860 & -52.0438 & GRB 200411A &38\\
 702355 &   48.4381 & -52.7810 & GRB 200411A &45\\
 702639 &   48.3066 & -52.7748   & GRB 200411A &167\\
 702999 &   48.1607 & -52.0628 & GRB 200411A &572\\
 705197 &   47.4973 & -52.3007 & GRB 200411A &47\\
 705414 &   47.4271 & -51.9934 & GRB 200411A &55\\
 707028 &   46.8995 & -51.7947 & GRB 200411A &218\\
 707094 &   46.8666 & -51.7357 & GRB 200411A &58\\
 708405 &   48.1628 & -51.5325 & GRB 200411A &62\\
 708781 &   47.5189 & -51.6759 & GRB 200411A &28\\
 709415 &   46.3704 & -52.3510 & GRB 200411A &22\\
\end{tabular}
\label{tab:unmatched}
\end{table}

\begin{figure*}
    \includegraphics[width=0.9\textwidth]{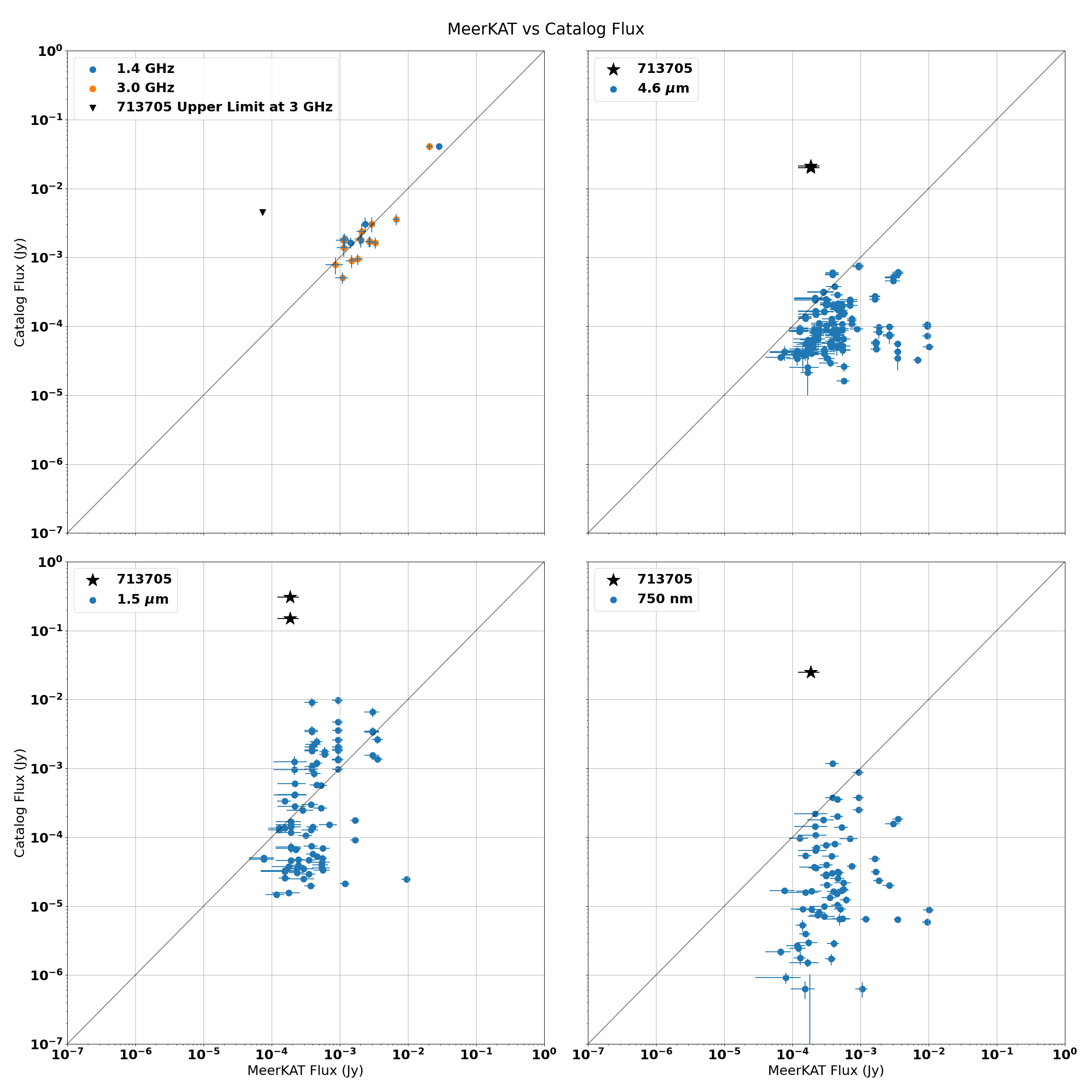}
    \caption{The average integrated flux measured in MeerKAT images on the horizontal axis and the catalog flux at various wavelengths from a variety of different catalogs are shown on the vertical axis. Source 713705 is a clear outlier and is highlighted with a black star symbol. The catalogs used are listed in the Acknowledgments section.}
\label{fig:combinedfluxfluxplots}
\end{figure*}

\begin{figure*}
\includegraphics[width=\textwidth]{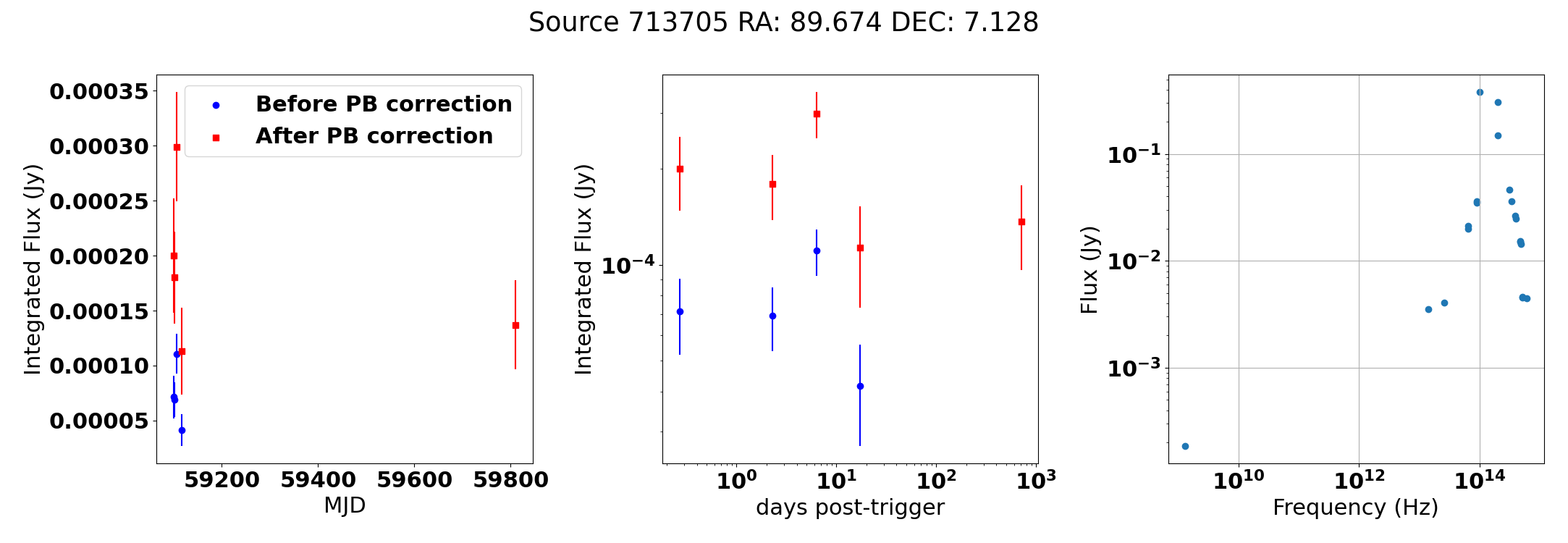}
\caption{Light curve and spectral energy distribution for source 713705, for which the optical counterpart is classified as a giant. The left panel shows the light curve on a linear scale, the middle panel the light curve on a log-log scale (with the start time being the trigger time of the GRB in the field), and the right panel is the spectral energy distribution with measurements from \citet{2003yCat.2246....0C,2012wise.rept....1C,2022yCat.1355....0G} as a part of the catalogs searched in this work.}
\label{fig:src713705lc4.png}
\end{figure*}

\begin{figure*}
	\includegraphics[width=\textwidth]{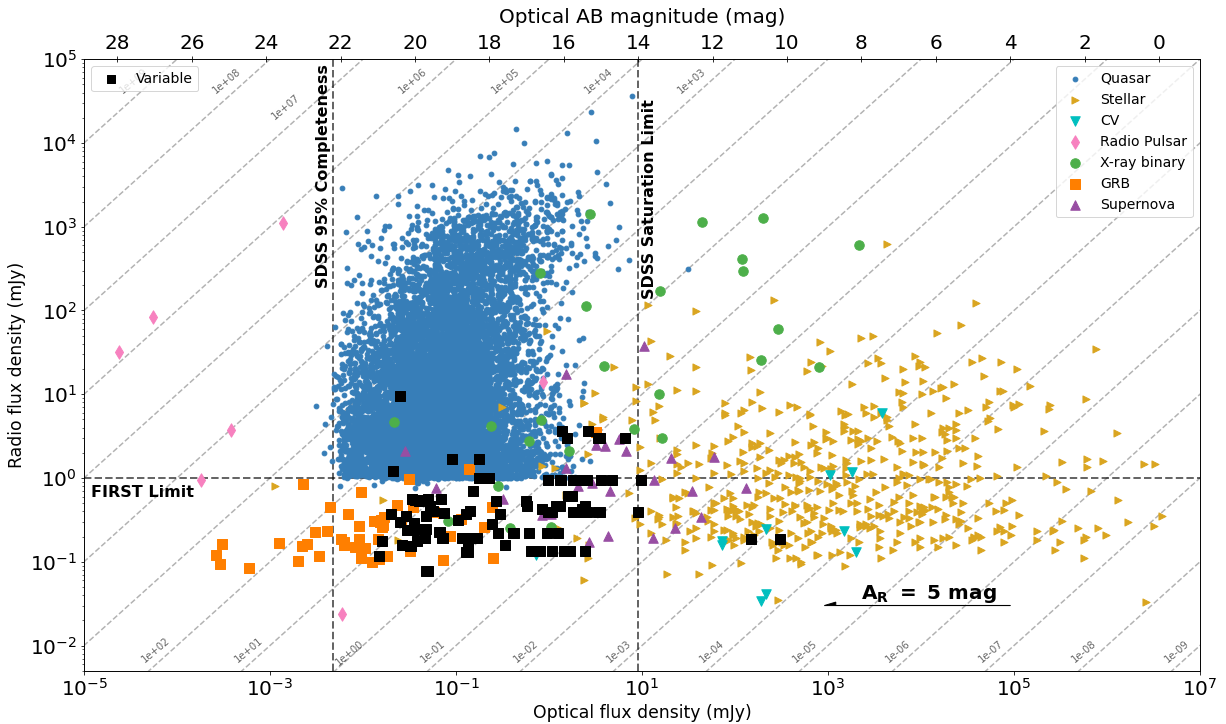}
    \caption{A plot from~\citet{2018MNRAS.479.2481S} showing radio vs optical (or near-infrared) flux density of a variety of variable sources with the fluxes of the catalog matched sources at $1500$ nm overplotted in black squares. Catalog matched fluxes are from~\citet{2003yCat.2246....0C,2006AJ....131.1163S,2013Msngr.154...35M,2007MNRAS.379.1599L}}
    \label{fig:stewartetalplot}
\end{figure*}

\subsection{Variable Source Characteristics}
Figure~\ref{fig:combinedfluxfluxplots} shows the average flux of our variable sources in the MeerKAT observations plotted on the horizontal axis and the associated catalog flux on the vertical axis. The matching catalog radio sources are close in flux to the averaged measured flux in MeerKAT. The proximity of the points to the 1:1 diagonal line shows how well they correspond. At other wavelengths, the flux does not appear to follow any specific correlation, but instead appears to be a cluster of sources with an outlier or two. These outliers are from a single source, source 713705 as identified in our TraP runs, that is classified as a star in multiple catalogs (light curve shown in Figure~\ref{fig:src713705lc4.png}). 

We also see this outlier source in Figure~\ref{fig:stewartetalplot}, where we have overplotted our variable sources with associations on the radio-optical classification from \citet{2018MNRAS.479.2481S}. From this figure, it is clear that our outlier is within the stellar sources, while the other sources are near the active galactic nuclei (AGN) and stellar explosions within this radio-optical parameter space. The other variables are very likely not supernovae or GRBs since they are variable sources over timescales that are characteristic of AGN and not supernovae or GRBs. Further catalog information about the outlier source from the TESS catalog version 8.2 gives the luminosity class of this source to be a giant. Two additional sources are classified as stars and have a luminosity class of dwarf. Eleven additional sources are classified as stars in the TESS 8.2 catalog, Guide Star Catalog 2.4.2, Dark Energy Survey Data Release 2, and Sloan Digital Sky Survey 16. In these same catalogs, 52 sources are either classified as extended sources or have multiple sources matched within the 1 arcsecond search radius.

\section{Discussion}
\label{sec:discussion}
Although there were no confidently detected transients, there remain a large number of variable sources that warrant further investigation. In addition, the lack of transient detections can be used to constrain parameter space via transient simulations. As part of further investigation of these sources, the nature of the variability, whether it is intrinsic or extrinsic variability, is examined here.

\begin{figure*}
    \includegraphics[width=\textwidth]{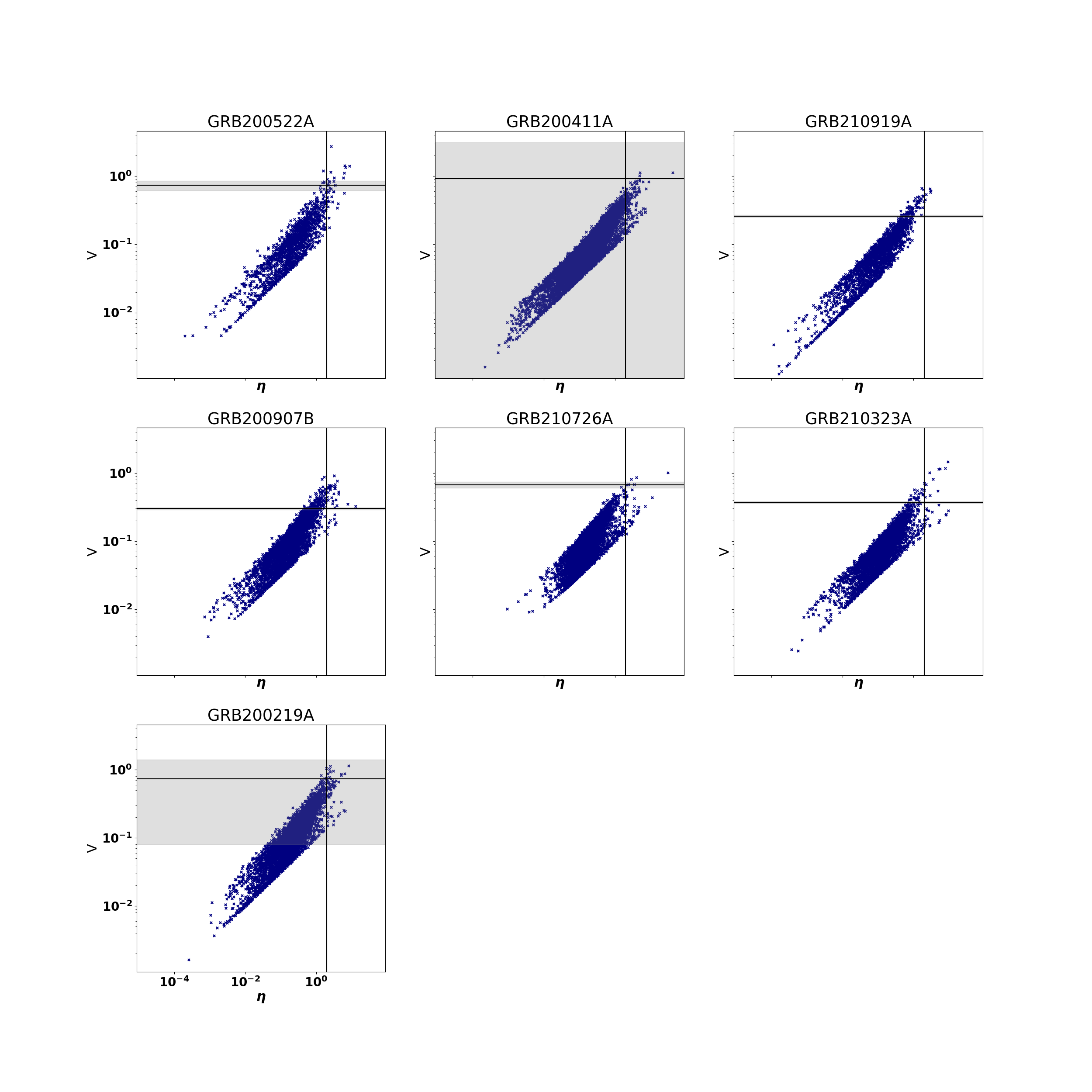}
    \caption{Scatter plot showing $V-\eta$ for each field with $\eta=2$ as a vertical line and the modulation index with errors from Table~\ref{tab:sciprop} shown as a horizontal line with a grey shaded region. The modulation index is defined similarly to the variability metric $V$ and is shown for comparison. The title of each subplot indicates the field that the sources are within.}
    \label{fig:allfieldsvar}
\end{figure*}

\begin{figure*}
    \includegraphics[width=\textwidth]{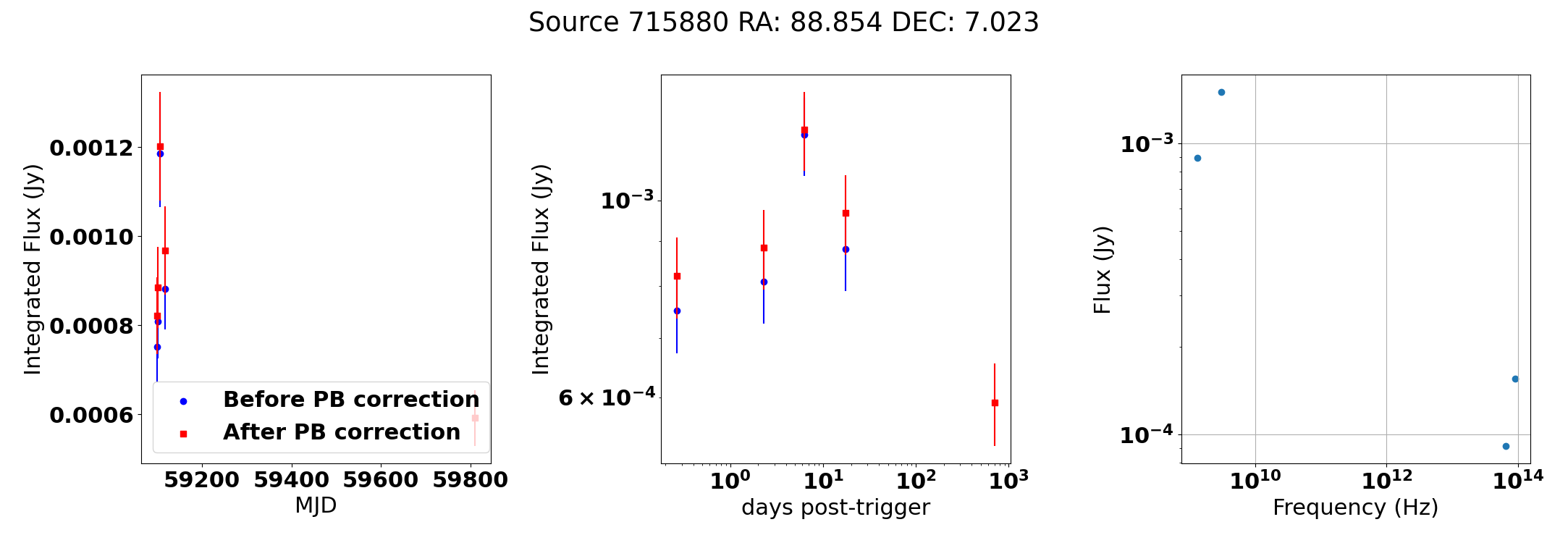}
    \caption{Light curve and spectral energy distribution for source 715880. The left panel shows the light curve on a linear scale, the middle panel the light curve on a log-log scale (with the start time being the trigger time of the GRB in the field), and the right panel is the spectral energy distribution from \citet{2008AJ....136..735L}  as a part of the catalogs searched in this work.}
    \label{fig:src715880lc4.png}
\end{figure*}

\subsection{Scintillation Effects}
\label{sec:scinteffects}
Interstellar scintillation is a known cause of variability in the radio sky. \citet{1998MNRAS.294..307W} provides some background on the kinds of variability expected from radio observations. Scintillation can occur when light at radio wavelengths interacts with inhomogeneities in the ionized component of the interstellar medium. The scattering produced by this interaction can be described as being either ``weak'' or ``strong.'' The dividing line between these regimes can be determined by comparing the observing frequency with the transition frequency $\nu_0$. When the observing frequency is approximately $\nu_0$, the modulation index, defined as $\sigma/\mu$ or the fractional variation in flux where $\mu$ is the average flux and $\sigma$ is the variation in flux, is equal to one. The timescale over which the modulation in flux, m, occurs is called the variability timescale, $t_{var}$. If the observing frequency is greater than the transition frequency, the observations are in the weak scattering regime, and if it is less than the transition frequency, the observations are in the strong scattering regime. Given the observing frequency of 1.3~GHz and typical $\nu_0$ values shown in Table~\ref{tab:sciprop}, all the observations in this survey are in the strong scattering regime. The strong scattering regime can be further broken down into refractive and diffractive scintillation. For all the timescales involved in these variable sources we are interested in examining refractive scintillation. ~\citet{2019arXiv190708395H} created models of refractive scintillation using $H_{\alpha}$ maps. Using these models and relations, we find that we expect scintillation to have a large effect on the amount and kinds of variability to expect in the light curves of individual sources.

\begin{table*}
\caption{Observed fields and some properties, together with the scintillation parameters calculated from~\citet{2019arXiv190708395H} for the center of each field. The parameters m and t$_{var}$ are defined in section~\ref{sec:scinteffects}}
\begin{tabular}{lrrrrll}
\hline
      Name & Average Noise  &  \# of variable sources &  Survey length  &  $\nu_0$ &             m & t$_{var}$  \\
       & ($\mu$Jy/beam) & & (days) & (GHz) & & (days)\\
\hline
GRB200219A &                   9 &                     17 &                  8.1 &      2.2 & $0.74\pm0.66$ &      $1.7\pm4.5$ \\
GRB200411A &                   7 &                     51 &                  6.2 &      1.5 & $0.92\pm2.22$ &      $0.9\pm6.5$ \\
GRB200522A &                  31 &                      4 &                 14.0 &      2.2 & $0.74\pm0.12$ &      $1.6\pm0.8$ \\
GRB200907B &                  12 &                     14 &                 17.2 &     10.9 &  $0.3\pm0.01$ &     $57.9\pm4.7$ \\
GRB210323A &                   9 &                     12 &                189.6 &      7.5 & $0.37\pm0.01$ &     $23.7\pm2.4$ \\
GRB210726A &                   9 &                     24 &                151.7 &      2.7 & $0.67\pm0.07$ &      $2.1\pm0.6$ \\
GRB210919A &                  18 &                      0 &                  7.2 &     14.2 & $0.26\pm0.01$ &     $62.4\pm4.1$ \\
\hline
\end{tabular}
\label{tab:sciprop}
\end{table*}

Table~\ref{tab:sciprop} shows a summary of all the fields in the survey, some of their properties, and the scintillation parameters calculated from~\citet{2019arXiv190708395H}. A closer look at this table may explain a large amount of the variability we see in our survey. For example, the field with the most transient detections, the GRB 200411A field, has a transition frequency $\nu_0$ that falls within the observing band. Consequently, the modulation index m is quite high for this field. Combining this information with the variability timescale $t_{var}$ reveals that in principle all the variables in this field can be explained by refractive scintillation. There are other fields in which $\nu_0$ is close to the observing band: using the same logic as for the GRB 200411A field, we can say that the variables in the GRB 200219A field, GRB 200522A field, GRB 210323A field, and GRB 210726A field can be explained by refractive scintillation. Note that the number of detected variables in the GRB 200522A field is lower due to the higher average noise. Any variable in the aforementioned fields would need to show a calculated modulation index greater than the already high expected modulation index from scintillation in these fields, and after examining these sources none of them have a modulation index significantly higher than that predicted for refractive scintillation. In Figure~\ref{fig:allfieldsvar}, we show how the modulation index, shown as a grey shaded region, compares to the variability parameter, V, in this scatter plot of $\eta$ and V for each field. This plot shows how for some fields, the modulation index is very high, and could be consistent with all of the sources in the field. Note, however, that this does not consider timescale of variability.

In the case of the GRB 210919A field, we see that the predicted scintillation timescale is much longer than the duration of the survey of this field. Therefore, it follows that no variables were detected. However, in the GRB200907B field, we also see a longer timescale for scintillation than the length of the survey, and in this field there are detected variable sources. All but two of the sources have variability timescales that are at least 17 days, as can be seen in source 715880 shown in Figure~\ref{fig:src715880lc4.png} and in source 713705 in Figure~\ref{fig:src713705lc4.png}. A possible explanation for this timescale and modulation index could be that the scattering screen is closer than is assumed in the estimates for scintillation. This possibility could be supported by the estimations on the refractive scintillation from~\citet{2019arXiv190708395H} showing one source that is very different in modulation index and timescale. This source ended up having variability more consistent with modulation indices and timescales like the rest of the field. Therefore, for this study, we took a single modulation index at the center of the field, but the point remains that there is a possibility that this region of the sky contains very inconsistent ISM charged particle populations that could possibly explain the variability.

\subsection{Intrinsic Variablity}
\begin{figure*}
    \includegraphics[width=0.8\textwidth]{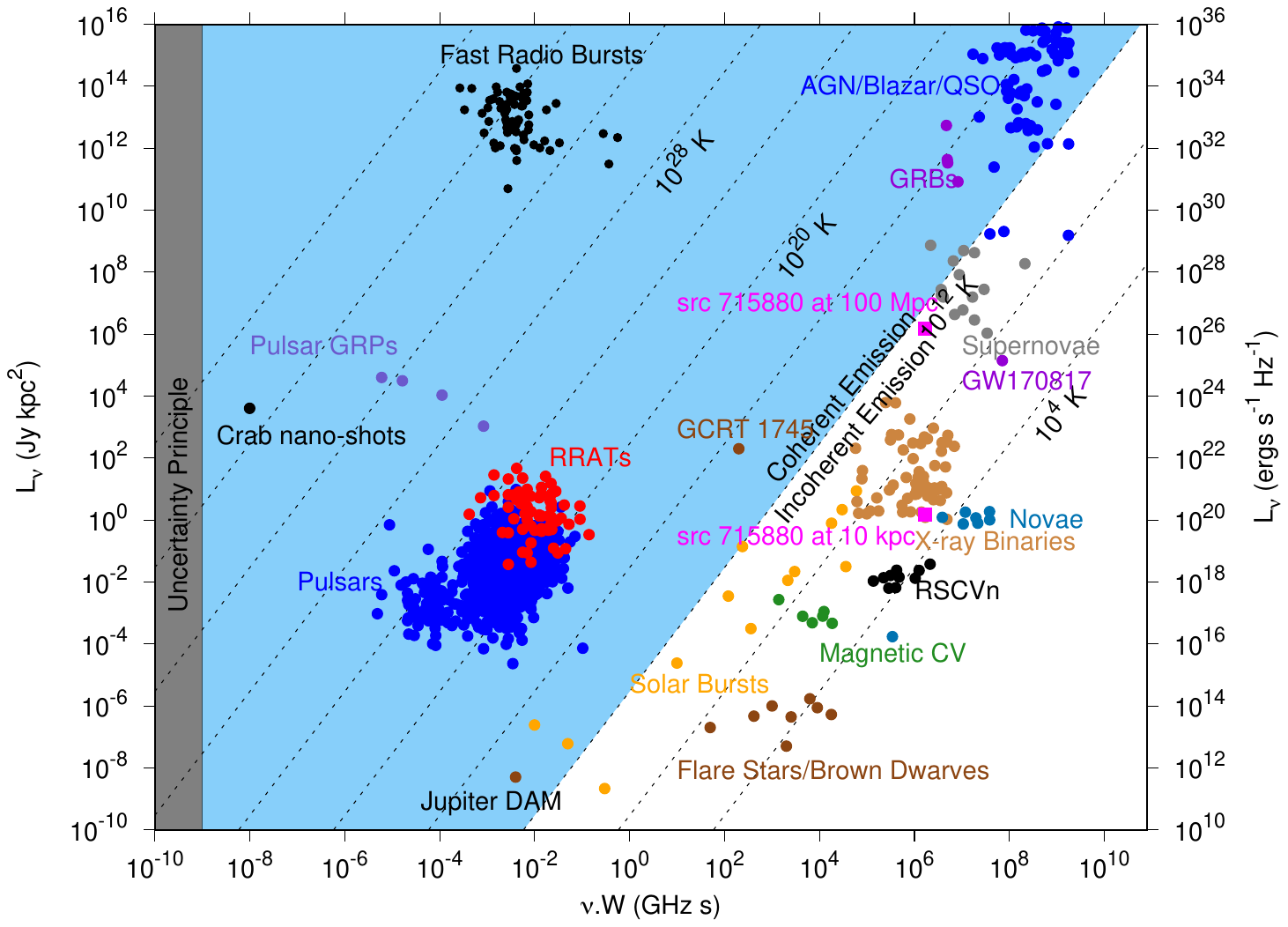}
    \caption{Scatter plot of a variety of transients observable in radio as a function of variability timescale and luminosity, adapted from~\citet{2015MNRAS.446.3687P}. Overplotted with pink squares are the values for source 715880 at both 10 kpc and 10 Mpc.}
    \label{fig:transphasespace}
\end{figure*}

Of the fourteen variable sources in the GRB 200907B field, two have variability on timescales shorter than 15 days, a timescale inconsistent with extrinsic variability from refractive scintillation according to the models we have used, therefore this variability is most likely intrinsic to these sources. Source 713705, also discussed in the previous section, is a known variable star also called ASASSN-V J055841.70+070741.7, with a period of 17.22 days reported in the AAVSO International Variable Star Index~\citep{2006SASS...25...47W}. This source is known to be variable at other wavelengths and could be intrinsically variable in the radio as well. The light curve for this source is shown in Figure~\ref{fig:src713705lc4.png}. The variability of this source is quite short with the flux rising to its peak and falling again within a timespan of about 15 days. Like source 713705, source 715880 shows a single high flux measurement in a fifteen day span which can be seen in Figure~\ref{fig:src715880lc4.png}. This source is not classified in any other catalogs, but has some flux measurements in the infrared with WISE~\citep{2014yCat.2328....0C}. Figure~\ref{fig:transphasespace} shows source 715880 with its variability timescale (assuming 15 days) multiplied by the observing frequency (1.3 GHz), against its luminosity. Two values are shown on the scatter plot for this source: one if the source is at 10 kpc and one if the source is at 10 Mpc. From this plot, we see that the source is possibly an X-ray binary or nova if it is at 10 kpc, and it is possibly a supernova if it is at 10 Mpc. Further investigation into this source, including follow-up observations, are warranted to classify the source and determine what sort of variability could cause its radio behaviour.

\begin{figure*}
    \includegraphics[width=\textwidth]{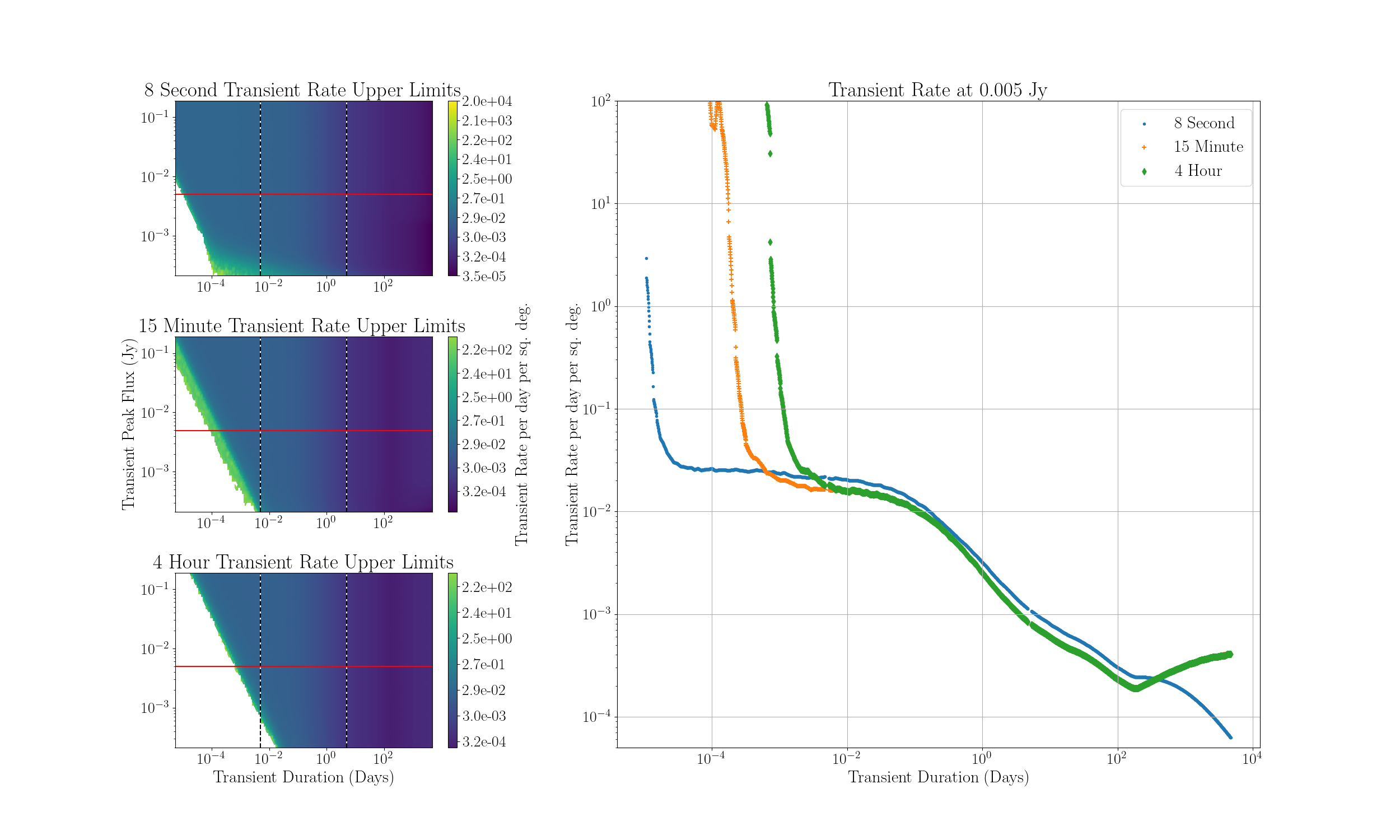}
    \caption{Transient rate upper limits for our survey based on the 8-second, 15-minute and 4-hour observations, calculated using the simulations code of ~\citet{2022ascl.soft04007C}. The left three panels show the transient rate upper limits color coded as a function of peak flux and duration. The panel on the right shows the transient rate as a function of duration at a given flux of 5 mJy, for the three different types of observations in our survey. Note that the dip downwards in upper limits on the 8 second timescale at long transient durations (approximately 100 days) is due to false transient detections and should be ignored.}
    \label{fig:threelimits}
\end{figure*}

\begin{figure*}
    \includegraphics[width=\textwidth]{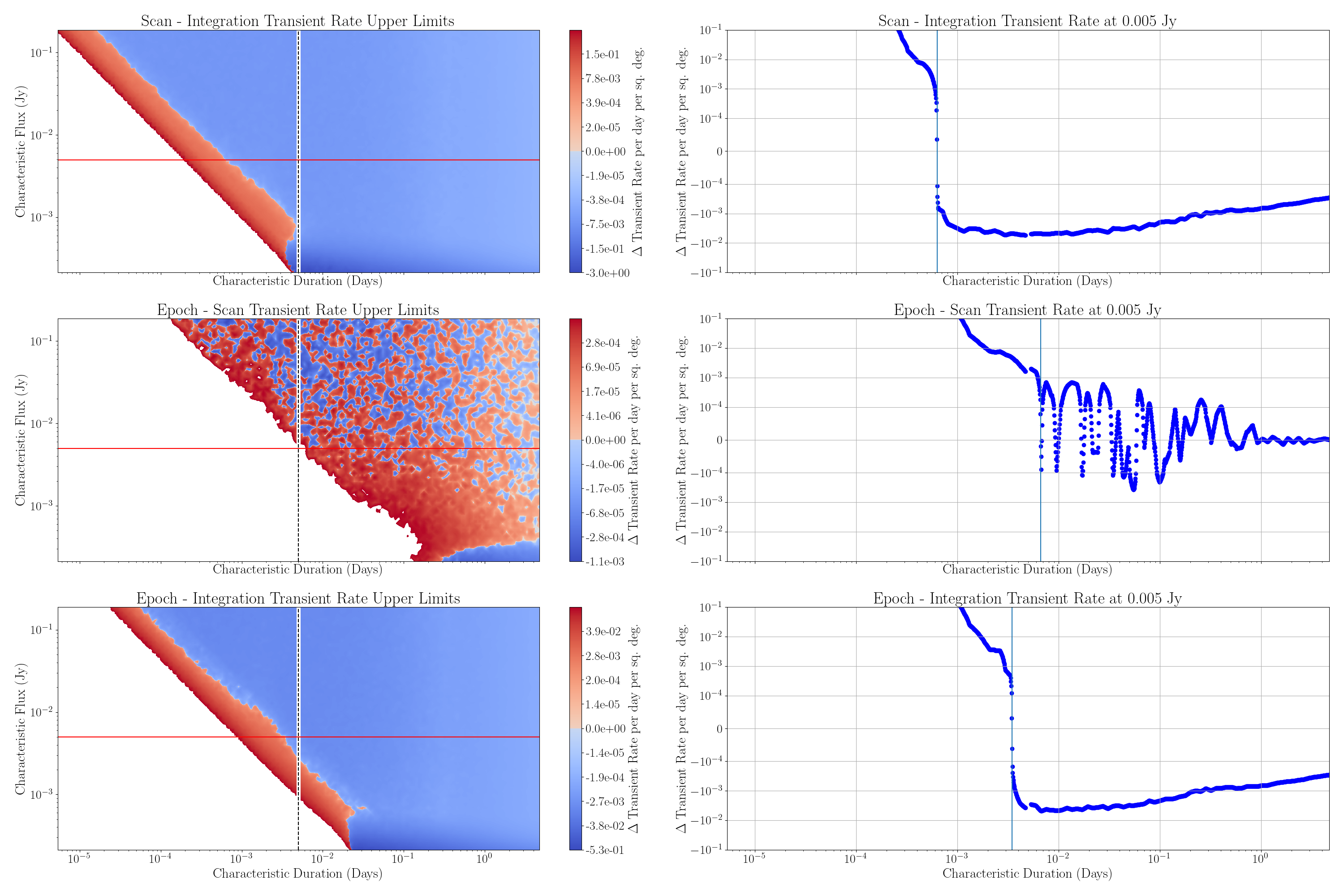}
    \caption{Difference in transient rate upper limits, calculated using the simulations code of ~\citet{2022ascl.soft04007C}, between the three different timescales of the survey. The left three panels show the differences between transient rate upper limits color coded as a function of flux and duration. The three right panels show this difference as a function of duration at a given flux of 5 mJy. The region that favors each timescale is noted by the corresponding color (red or blue) and the sign (positive or negative).}
    \label{fig:threediff}
\end{figure*}

\subsection{Transient Rate Limits}
Calculating transient rates is a key way to characterize transients and compare the sensitivity of surveys like this one to other surveys in various parts of transient parameter space. Transient rates are often calculated by assuming that transients are distributed as Poisson distributions and the rate is calculated by determining the number of detections over the duration of the survey. However, as discussed in~\citet{2017MNRAS.465.4106C} and~\citet{2016MNRAS.459.3161C}, many observational effects are often ignored, such as gaps within a survey or within individual observations, which leads to estimated transient rates that are off by orders of magnitude. Therefore, in order to place accurate limits on the transient rate imposed by the survey presented here, we use the transient simulations~\citep{2022ascl.soft04007C} as described in detail in~\citet{2022A&C....4000629C}. In summary, by using Monte-Carlo simulations, these simulations read in the metadata from the survey such as the observation times, times on target during the observations, locations, fields of view, image noise, and other inputs, and generate a large number of simulated sources, testing to see which sources would be detected. The results are binned in transient flux and duration, and probabilities are generated. These probabilities are used to compute transient rates by assuming transients follow a Poisson distribution.

In Figure~\ref{fig:threelimits}, the transient rate upper limits are shown for the 8-second, 15-minute, and 4-hour images. The left panels show the transient duration on the horizontal axis, transient peak flux on the vertical axis, and the transient rate upper limit in the color axis. The right panel shows the transient duration on the horizontal axis and the vertical axis shows the calculated transient rate for a transient with a flux density of 5 mJy. Because these timescales all probe the same field, we can show them all in the same panel on the right by just taking the strictest upper limits on transient rates from each timescale. Note that the dip downwards in upper limits on the 8 second timescale at long transient durations (approximately 100 days) is due to false transient detections and should be ignored. These types of false detections can be mitigated by looking for the long timescale transients in the deeper images.

We further examined which parts of transient parameter space are best probed by which timescale. Figure~\ref{fig:threediff} shows the difference in calculated transient rate for all combinations of the timescales. The panels on the left show the transient duration on the horizontal axis, the transient flux on the vertical axis, and the difference between the transient rate upper limits on the color axis. Since lower limits are better, the timescale with the lowest limits are noted with either a red or blue color corresponding to either positive or negative values of the equation noted in the title of each plot. The panels on the right side show the difference in transient rate on the vertical axis for a transient at a flux of 5 mJy with the duration on the horizontal axis. The top plots show the difference between the 8-second and 15-minute timescales; the middle plots show the difference between the 4-hour and 15-minute timescales, and the bottom plots show the difference between the 4-hour and 8-second timescales. The top and bottom plots appear to show a certain fluence where the timescale that gives the lower limits changes. The middle plot shows less of a difference between the two timescales, although judging from the transient durations where each timescale gives lower limits in the top and bottom plots, there appears to be a small region in which the transient rate upper limit is lowest in the scans. 

\begin{figure*}
    \includegraphics[width=\textwidth]{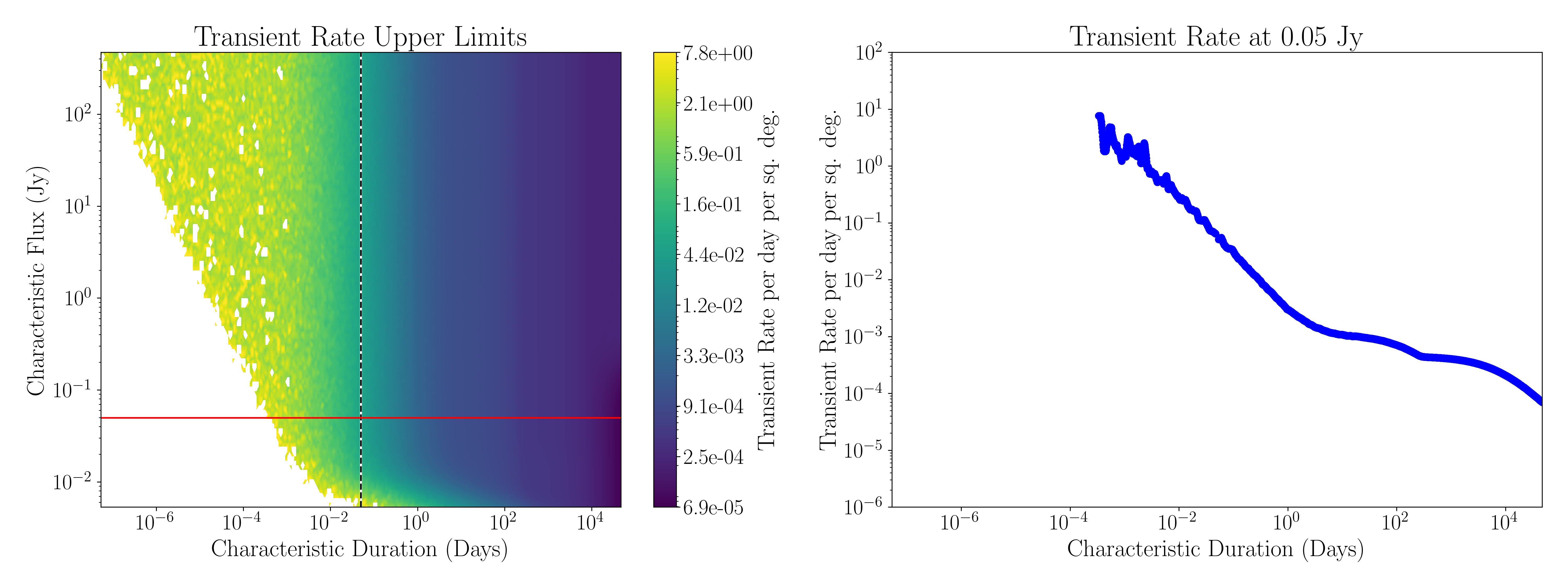}
    \caption{Transient rate limits for a survey similar to~\citet{2011ApJ...728L..14B}. The vertical dashed line in the left panel marks where two different simulations were combined into a single plot; the horizontal red line marks 50 mJy. The panel on the right shows the transient rate as a function of duration at 50 mJy.}
    \label{fig:bowerandsaulall}
\end{figure*}

In order to compare our results with other surveys, we also calculated transient rates using this method for a survey similar to~\citet{2011ApJ...728L..14B}, in which a commensal transient search was performed on archival calibrator observations of 3C 286 spanning 23 years on a cadence that is approximately weekly or slightly better than weekly. We use the same sensitivity in our simulations as is used in their survey. For the observation dates, we only have the information on the day and no information on the duration, so we take the time to be at midnight and set the duration to be sometime between 1.75 minutes and 2.25 minutes. We set the field of view to be 1 degree across. Using this setup, we create Figure~\ref{fig:bowerandsaulall}, which we can compare to our survey results in Figure~\ref{fig:threelimits}. The transient flux that our survey is sensitive to is at least an order of magnitude deeper, and in the case of the 4-hour timescale even two orders of magnitude deeper. However, a survey like~\citet{2011ApJ...728L..14B} has a deeper transient rate for higher flux and longer timescale transients. The transient rate upper limits of the latter are particularly constraining for transients with a duration of over 1000 days for a transient with flux of 50 mJy. In comparison, our survey has transient rate upper limits that are relatively flat and constraining all the way down to $2\times10^{-5}$ days, i.e., a few seconds, at a transient flux of 5 mJy. 

\subsection{Comparing Limits on FRBs}
We examined the possibility of detecting an FRB in our survey. FRBs have timescales that are much shorter than the timescale of our observations, down to milliseconds. However, the flux of these sources is also quite high, exceeding Jansky levels. For this reason, it may be possible to detect an FRB in the 8 second images since the total fluence may be sufficient for it to be detected. Note that in this study we are limited to full bandwidth observations and may miss a burst that is relatively narrow-band. Future surveys could present opportunities to search for narrow-band transient events by splitting up the observations in frequency.
 For example \citet{2023MNRAS.518.3462A}, detected an FRB in two second images and would have been able to detect the brightest bursts in eight second images.  \citet{2021ApJS..257...59C} calculates the rate of FRBs based on CHIME observations to be $820\pm 60 (\text{stat.})_{-200}^{+220}(\text{sys.})$ per sky per day. If we convert this value to a transient rate per square degree we find: $1.99\times10^{-2}\pm0.15_{-0.49}^{+0.53}$. The fluence in conjunction with the integration time of the observations is what determines what is detectable in a survey. The minimum fluence that we can detect is approximately 10 Jy ms. The transient rate at this fluence is approximately $2.5\times10^{-2}$ transients per day per square degree.  If we follow the scaling outlined in~\citet{2021ApJS..257...59C}, we can use $\alpha=-1.4$, where $\alpha$ is the power-law index for the scaling of FRB-like sources above a certain fluence, to rescale the fluence we detect to compare it with CHIME. From this we find a modified upper limit of approximately $6.6\times10^{-2}$  possible FRB-like transients per day per square degree. 
 
\begin{figure*}
    \includegraphics[width=\textwidth]{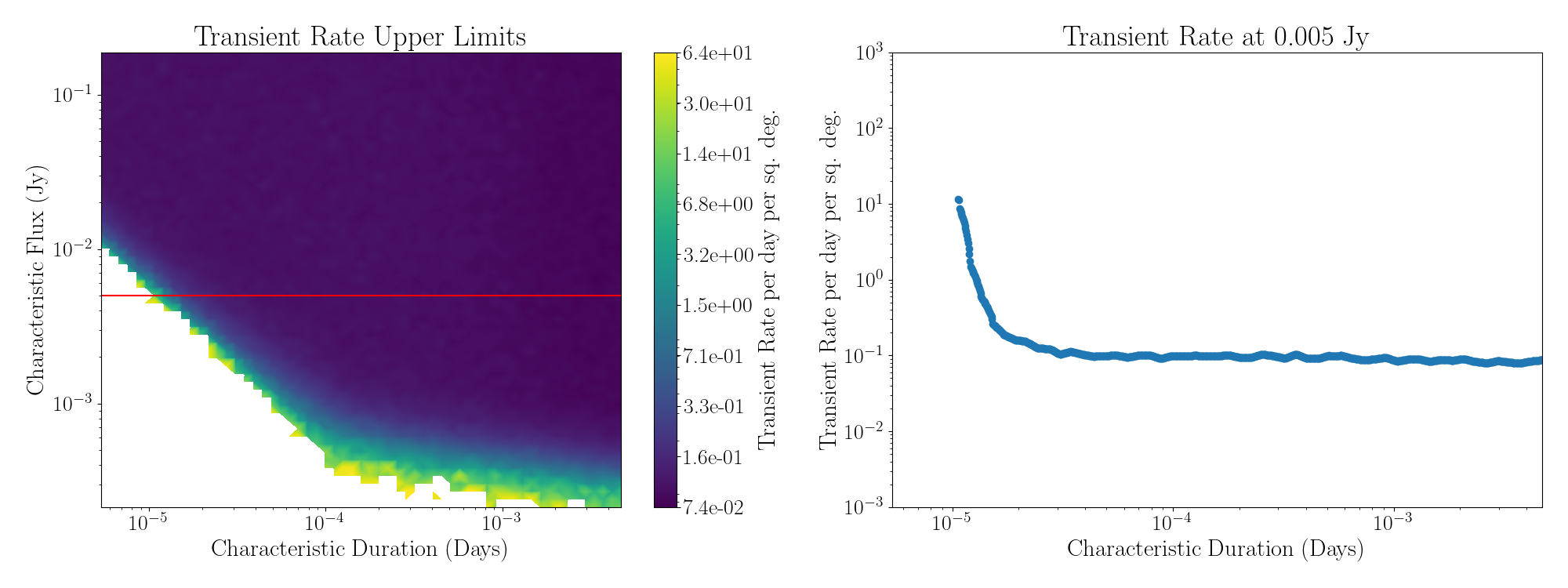}
    \caption{Transient rate upper limits, calculated using~\citet{2022ascl.soft04007C}, for a survey exactly double in duration of the 8-second timescale survey.}
    \label{fig:doublerate}
\end{figure*}

 By comparison, the minimum fluence of a survey like~\citet{2011ApJ...728L..14B} is around 1 kJy ms. Our survey does not yet give limits on transient rates that would be below that of the CHIME FRB rates. In order to see the feasibility of discovering an FRB-like event in a survey like ours, we simulate the same 8-second timescale images in our survey but with the observation continuing for another two years with a similar set-up. The resulting transient rates can be seen in Figure~\ref{fig:doublerate}. From this figure we see that the upper limit on the transient rate would be around $10^{-1}$ transients per day per square degree at the same minimum fluence as before. Rescaling the fluence results in an upper limit of $2.64\times 10^{-1}$ possible FRB events per day per square degree. This means that with a doubling of the survey length, we still will not quite be able to place tighter limits on the FRB population. However, since this search is a commensal search and these limits are approaching the rates set by CHIME, it is worthwhile to continue to search for these short timescale transients in order to refine our understanding of them. 

\section{Conclusion}
We search deep MeerKAT observations of short gamma-ray burst fields for transients and variable sources, by making images with integration times of 4 hours, 15 minutes, and 8 seconds. This results in a transient survey that spans timescales from seconds to months. We search for transients in these images using the LOFAR Transients Pipeline~\citep{2015A&C....11...25S}. While we do not find any significant transients in the 8-second and 15-minute images, we find more than 120 variables in the long observations. Most of the variability can be explained by interstellar scintillation effects on the radio emission from active galactic nuclei. However, in a few cases the variability is likely intrinsic, because the observed modulation and variability timescales differ significantly from expectations and the variability observed in other sources in those fields. We also place new, accurate limits on the transient rate using transient simulations~\citep{2022ascl.soft04007C}. Our limits at the shortest timescales and lowest fluence levels are approaching the limits placed by time series searches at sub-second timescales such as those for FRBs with CHIME \citep{2021ApJS..257...59C}. Continued commensal searches, in conjunction with refining the techniques for transient searches that are described here, should continue to constrain transient rates calculated from image searches and approach the rates found for sources such as FRBs, thus providing a new method for studying transients on short timescales.

\section{Acknowledgments}
We would like to thank Geoff Bower for sharing information and data from \citet{2011ApJ...728L..14B}. 

The MeerKAT telescope is operated by the South African Radio Astronomy Observatory (SARAO), which is a facility of the National Research Foundation, an agency of the Department of Science and Innovation. We would like to thank the operators, SARAO staff and ThunderKAT Large Survey Project team.

We acknowledge use of the Inter-University Institute for Data Intensive Astronomy (IDIA) data intensive research cloud for data processing. IDIA is a South African university partnership involving the University of Cape Town, the University of Pretoria and the University of the Western Cape. We also acknowledge the computing resources provided on the High Performance Computing Cluster operated by Research Technology Services at the George Washington University.

This work was completed in part with resources provided by the High Performance Computing Cluster at The George Washington University, Information Technology, Research Technology Services.

This research has made use of the VizieR catalogue access tool, CDS,
 Strasbourg, France (DOI : 10.26093/cds/vizier). The original description 
 of the VizieR service was published in 2000, A\&AS 143, 23

A. Andersson acknowledges the support given by the Science and Technology Facilities Council through a STFC studentship. 

We would like to thank Ben Stappers for his feedback. Additionally, we would like to thank Oleg Kargaltsev and Bethany Cobb-Kung for their helpful comments.

We would also like to thank the referee for the valuable comments that improved the quality of the paper. 

The catalogs that were searched for multi-wavelength counterparts and included in Figure~\ref{fig:combinedfluxfluxplots} are~\citet{vizier:V/156,vizier:V/154,vizier:J/ApJS/249/18,vizier:J/ApJ/794/120,vizier:J/A+A/540/A106,vizier:II/371,vizier:II/358,vizier:II/349,2006AJ....131.1163S,vizier:II/367,vizier:II/319,vizier:II/246,vizier:B/denis,vizier:J/ApJS/225/1,vizier:II/363,vizier:II/328,vizier:II/311,1998AJ....115.1693C,vizier:VIII/92}.

\section*{Data Availability}

Data is available upon request by email to schastain@gwu.edu.


\bibliographystyle{mnras}
\bibliography{inprogress} 





\bsp	
\label{lastpage}
\end{document}